\documentclass[aip,apl,reprint]{revtex4-1}
\usepackage[dvips]{graphicx}

\begin{document}

\bibliographystyle{apsrev}

\title{Femtosecond Pump-Probe Studies of Reduced Graphene Oxide Thin Films}

\author{Brian A. Ruzicka}

\author{Lalani K. Werake}

\author{Hui~Zhao}\email{huizhao@ku.edu}

\affiliation{Department of Physics and Astronomy, The University of Kansas, Lawrence, Kansas 66045}

\author{Shuai Wang}

\author{Kian Ping Loh}

\affiliation{Department of Chemistry, National University of Singapore, 3 Science Drive 3,Singapore 117543}

\begin{abstract}
The dynamics of photocarriers in reduced graphene oxide thin films is studied by using ultrafast pump-probe spectroscopy.  Time dependent differential transmissions are measured with sample temperatures ranging from 9 to 300 K. At each sample temperature and probe delay, the sign of differential transmission remains positive. A fast energy relaxation of hot carriers is observed, and is found to be independent of sample temperature. Our experiments show that the carrier dynamics in reduced graphene oxide is similar to other types of graphene, and that the differential transmission is caused by phase-state filling of carriers.
\end{abstract}


\maketitle

Graphene consists of a single two dimensional atomic layer of carbon atoms arranged in a hexagonal lattice. \cite{s306666}  The popularity of graphene has soared recently in part due to its superior properties, which include extremely high thermal conductivity \cite{nl8902}, high intrinsic breaking strength \cite{s321385}, and very high charge carrier mobility. \cite{s306666}  Currently, there are four main types of graphene:  graphene produced by mechanical exfoliation of bulk graphite,\cite{s306666} epitaxial grown graphene, \cite{s3121191} graphene produced by chemical vapor deposition \cite{s3241312,n457706}, and chemically derived graphene. \cite{nl1092,nn3270}

Recently, the dynamics of photocarriers in graphene has been studied by ultrafast pump-probe techniques. \cite{apl92042116,apl95191911,nl84248,oe172326,b80245415}  These studies have focused on samples of epitaxial graphene, \cite{apl92042116,l101157402,nl84248,apl94172102,b80245415,apl96081917} graphene produced by the micromechanical cleavage method, \cite{oe172326} and very recently graphene produced by chemical vapor deposition.\cite{apl96081917} While these three types of graphene provide great opportunities to study the physics of graphene, chemically derived graphene not only shares this quality, but has also shown great promise for effective use in industry.  Simple and reproducible methods have been developed to create large graphene films from graphene oxide.  \cite{nn3270}  Also, high mobility printable graphene films have been developed, which can be of great use in a graphene-based field-effect transistor. \cite{nl1092}  Therefore, in order to develop graphene related applications in the future, it is important to study carrier dynamics in this type of graphene.  In addition to this, optical studies of chemically derived graphene can be very useful because chemically derived graphene can be placed on any substrate, which allows for the ability to ensure that the substrate plays no role in the experimental results.  This is different from optical studies of epitaxial graphene grown on SiC, where the first few layers are doped due to the SiC, and optical studies of graphene produced by mechanical exfoliation, which require a substrate of SiO$_{2}$/Si to locate the graphene layers. \cite{apl92042116,nm6770,oe172326}  Very recently, ultrafast carrier dynamics in reduced graphene oxide in different solutions was studied by Kumar {\it et al}. \cite{apl95191911}  However, reduced graphene oxide in solution form also has certain limitations.  For example, it does not allow for the possibility to study how the carrier dynamics change with a varying lattice temperature.

Here, we present a study of the carrier dynamics of reduced graphene oxide thin films using ultrafast pump-probe spectroscopy. The time dependent differential transmission is measured with sample temperatures ranging from 9 to 300~K. We found that at all sample temperatures and probe delays, the sign of the differential transmission remains positive. The fast energy relaxation of hot carriers is observed, and is found to be independent of sample temperature. Also, our experiments show that the carrier dynamics in reduced graphene oxide is similar to other types of graphene, and the differential transmission is caused by phase-state filling of carriers.

\begin{figure}
 \includegraphics[width=7.5cm]{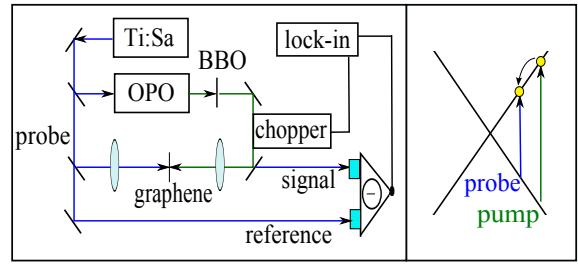}
 \caption{Experimental setup and excitation/probe scheme.}
\end{figure}

In our experiment, graphene oxide flakes, synthesized using a modified Hummers' method, \cite{jacs801339} were spin coated on a quartz substrate to form a film, which was then transformed to a multi-layer graphene film by thermal reduction at 1000$^\circ \mathrm{C}$. \cite{nl1092} Since a thick sample is desirable for optical transmission measurements, a high concentration solution (4mg/ml) was used and the spin-coating was repeated for four times. The number of graphene layers is determined to be about 50 by using an atomic force microscope. The absorption of the sample at 750-nm is measured to be about 50$\%$.

Figure 1 summarizes the experimental setup (left) and excitation/probe scheme (right).  Carriers are first excited with a pump pulse with a central wavelength of 750 nm, a pulse width of 0.1 ps, and a spot size of approximately $2.3$ $\mu$m at full width at half maximum (FWHM). The pump pulse is obtained from second-harmonic generation using a beta barium borate (BBO) crystal of the signal output of an optical parametric oscillator (OPO) pumped by a Ti:sapphire laser (Ti:Sa) at 80 MHz. To detect the carrier dynamics, we use a probe pulse with a central wavelength of 810 nm, a spot size of $1.2$ $\mu$m (FWHM), and a pulse width of 0.19 ps.  When carriers occupy the probing energy states (0.765 eV, half of the probe photon energy), the transmission of the probe through the sample will increase due to state filling effects.  Hence, by detecting the differential transmission $\Delta T / T_{0} \equiv [T(N)-T_{0}]/T_{0}$, i.e. the normalized difference in transmission of the probe with $[T(N)]$ and without ($T_{0}$) carriers at different probe delays, we can deduce the density of carriers in these states ($N$). The differential transmission is measured by sending a portion of the transmitted probe pulse to one photodiode of a balanced detector, which is connected to a lock-in amplifier referenced to an optical chopper that is modulating the intensity of the pump beam.  In order to decrease noise caused by the intensity noise of the probe pulse, a reference pulse is sent to the other photodiode of the balanced detector.\cite{l96246601}

\begin{figure}
 \includegraphics[width=7.5cm]{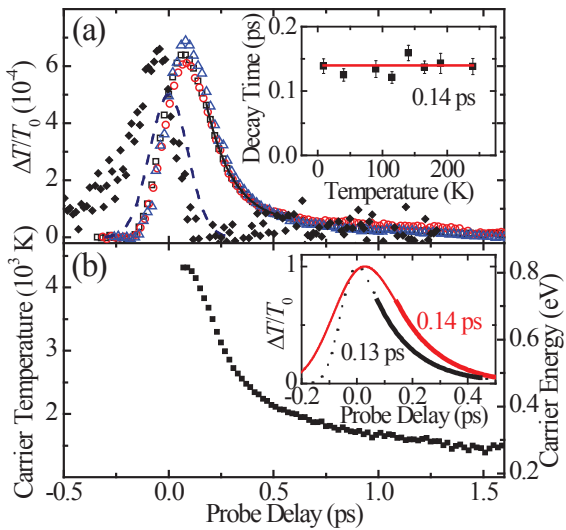}
 \caption{(a) $\Delta T / T_{0}$ measured for sample temperatures of 9 (squares), 115 (circles), and 240 K (up-triangles). The diamonds show a differential transmission measured with the same setup but with the 810-nm pulse as pump and the 750-nm pulse as probe, that is scaled to the same magnitude of the other curves. The dashed line shows the cross-correlation of the pump and the probe pulses, calculated using the known pulse widths. The inset shows the decay time of the differential transmission signal for various lattice temperatures. (b) Carrier temperature (left axis) and average carrier energy (right axis) deduced from the $\Delta T / T_{0}$ curve at 9 K. The inset shows the de-convolution of the probe pulse.}
\end{figure}

The time dependent differential transmission is measured for lattice temperatures ranging from 9 to 300 K, with a few examples shown as the symbols in Fig.~2(a). In these measurements, the peak energy fluence of the pump pulse is 118~$\mu$J cm$^{-2}$. Using the absorption of 50$\%$, the excited areal carrier densities in the first and the last graphene layers are estimated to be about  $8.0 \times 10^{12}$ and $3.3 \times 10^{12}$ cm$^{-2}$, respectively. We note that although the density in the front layer is about 20$\%$ of the density of state in graphene within the pump bandwidth (0.02 eV),\cite{pr71622} no significant absorption saturation is expected due to ultrafast thermalization of carriers: the carriers excited by the earlier part of the pump pulse are rapidly scattering to other energy states, making room for the excitation of carriers by the latter part of the pulse. In order to determine the zero probe delay, defined as the probe delay time at which the centers of the pump and probe pulses overlap, we perform a differential transmission measurement in the same setup but with the 750-nm pulse as the probe and the 810-nm pulse as the pump (diamonds, scaled to match the height of the other three curves). The difference in the shape of these curves is caused by the fact that the delay time is defined as the time lag of the 810-nm pulse with respect to the 750-nm pulse. We define the crossing time of this reversed differential transmission curve with the other three curves as the zero probe delay. Given that the duration of the dynamics is so short, we plot in Fig.~2(a) the cross-correlation of the pump and probe pulses calculated from the known temporal width of the two pulses (dashed line). Clearly, the carrier dynamics persist longer than the cross-correlation.

The results are consistent with the following picture of the carrier dynamics that has been established in previous studies of graphene: After the carriers are excited with the pump pulse, they quickly reach a hot distribution via carrier-carrier scattering within a time scale on the order of 0.1 ps.  \cite{apl92042116,l101157402,nl84248,oe172326,apl95191911} Then, the carriers cool through carrier-phonon scattering (mainly optical phonon emission) on a slower time scale on the order of 1 ps.\cite{apl96081917}  This energy relaxation causes the decrease of the carrier density in the probe state, and therefore the decrease of $\Delta T / T_{0}$. The carrier recombination occurs on a much longer time scale, and is not seen in our experiment with rather high probe energy.

To investigate the effects of the lattice temperature on the relaxation of the carriers, a time constant for the decay of the differential transmission signal was determined by fitting a portion of the differential transmission curve (from 0.2 to 0.6 ps) with a single exponential decay function. The solid line in Fig.~2(a) shows an example. The obtained decay time constants are plotted in the inset of Fig.~2(a). No systematic variation is observed, and an average value of 0.14~ps is obtained. The inset of Fig.~2(b) shows the influence of the finite probe pulsewidth on the measured relaxation time. The dots show a simulated differential transmission signal assuming the probe pulse is infinitely short. An exponential fit to the corresponding section (solid line over the dots) gives a decay time of 0.13 ps. By convoluting such a curve with a 0.19-ps probe pulse, we obtain the thin solid line, where an exponential fit (thick solid line) gives a decay time of 0.14 ps. Therefore, we conclude that the measured relaxation time of 0.14 ps corresponds to a true relaxation time of 0.13 ps.

The fact that there is no dependence of the carrier energy relaxation on the lattice temperature is, however, not surprising.  As stated previously, immediately after excitation, the carriers form a hot distribution. The subsequent energy relaxation is dominated by optical phonon emission. Due to the high excitation excess energy, a large number of optical phonons is emitted, which causes a significant deviation of the phonon distribution in the excitation spot away from the equilibrium distribution dictated by the sample temperature. Since the carriers experience a much higher lattice temperature than the temperature at other parts of the sample, the sample temperature has no influence on the carrier dynamics.

Assuming a Fermi-Dirac distribution is rapidly established right after the excitation, \cite{apl92042116,apl94172102,nl84248,b80245415,apl96081917} we estimate that the initial temperature of the distribution is approximately 4,300 K.  Since $\Delta T / T_{0}$ is proportional to the density of carriers at the probing energy, we can calculate how the temperature and the average energy of the carriers changes over time using the measured $\Delta T / T_{0}$.  These calculations are shown in Fig.~2(b) using data from the 9 K measurement -- the calculations for measurements taken with different lattice temperatures give roughly the same values and show the same behavior.  The temperature of the carriers decreases to approximately 1,500 K within 1.5 ps, and then decreases much more slowly after this.

Interestingly, in previous low temperature experiments performed on epitaxial graphene, the differential transmission of the probe is actually negative for certain probe delays in the range of 0.3 to 1 ps,\cite{b80245415,l101157402} while in this experiment, the differential transmission is always positive for all ranges of lattice temperatures.  Plochocka {\it et al} attribute the negative differential transmission to an indirect absorption of the probe caused by many-body processes.\cite{b80245415}  Since we do not see the same behavior, we can't confirm this interpretation.  Instead, the probing mechanism in our sample is interband absorption throughout the entire relaxation process, and many-body effects do not play a role in changing the absorption of the probe.  We should also note that Sun {\it et al} attribute the decrease in probe transmission to an effect caused by the doped layers of graphene present due to the SiC substrate, which can only be seen for lower probe energies.\cite{l101157402}

\begin{figure}
 \includegraphics[width=7.5cm]{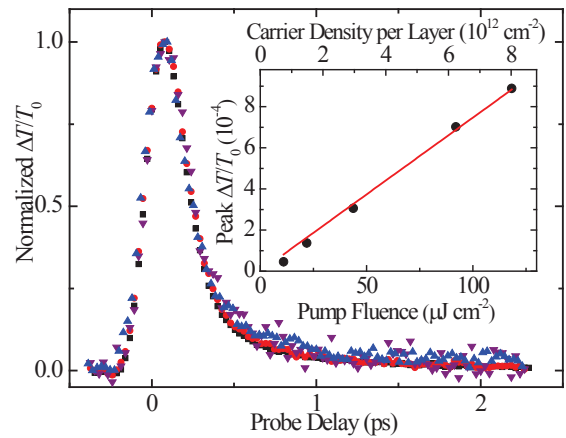}
 \caption{Normalized $\Delta T/T_{0}$ for pump fluences of 22 (squares), 44 (circles), 92 (up-triangles), and 118 $\mu$J/cm$^{2}$ (down-triangles) for a fixed sample temperature of 293 K.  The inset shows the peak $\Delta T/T_{0}$ for various pump fluences.}
\end{figure}

Finally, we study the carrier dynamics for various pump fluences, and hence various carrier densities.  Figure 3 shows the normalized $\Delta T/T_{0}$ for several pump fluences at room temperature.  Clearly, the energy relaxation is not affected by carrier density.  Also, the peak $\Delta T / T_{0}$ was measured for various pump fluences, as shown in the inset of Fig.~3.  We observe that the the peak $\Delta T / T_{0}$ is proportional to the pump fluence (solid line).  This confirms that the change in absorption of the probe is due to state filling effects and that the differential transmission is in fact proportional to the carrier density.  Also, we note that the lattice temperature did not affect this result.  As seen in Fig.~2(a), the peak differential transmission did not change significantly over the range of lattice temperatures studied when the pump fluence was fixed.

In summary, we have studied the carrier dynamics of reduced graphene oxide thin films using ultrafast pump-probe spectroscopy. Time dependent differential transmissions are measured with sample temperatures ranging from 9 to 300~K. We found that at all sample temperatures and probe delays, the sign of differential transmission remains positive. A fast energy relaxation of hot carriers is observed, and is found to be independent of sample temperature. Also, the differential transmission is proportional to the carrier density. Our experiments show that the carrier dynamics in reduced graphene oxide is similar to other types of graphene, and the differential transmission is caused by phase-state filling effects of carriers.

\begin{acknowledgements}
HZ acknowledges support from the US National Science Foundation under Awards No. DMR-0954486 and No. EPS-0903806, and matching support from the State of Kansas through Kansas Technology Enterprise Corporation. KPL thanks the support of NRF-CRP "Graphene Related Materials and Devices" (Grant No. R-143-000-360-281).
\end{acknowledgements}

\end{document}